\documentclass[11pt]{article}

\usepackage[margin=1in]{geometry}
\usepackage[T1]{fontenc}
\usepackage{lmodern}
\usepackage{microtype}
\usepackage{amsmath,amssymb}
\usepackage{graphicx}
\usepackage{booktabs}
\usepackage{xcolor}
\usepackage{algorithm}
\usepackage[noend]{algpseudocode}
\usepackage{enumitem}
\usepackage[numbers,sort&compress]{natbib}
\usepackage[colorlinks=true,linkcolor=blue!60!black,citecolor=blue!60!black,urlcolor=blue!60!black]{hyperref}
\usepackage[capitalise,noabbrev]{cleveref}

\newif\ifshowtodos
\showtodostrue

\newcommand{\sys}{INC-MoE}
\newcommand{\AG}{\textsc{AG}}
\newcommand{\RS}{\textsc{RS}}

\title{INC-MoE: Overlapping In-Network Collectives and Expert Computation for Expert-Parallel Mixture-of-Experts Inference}

\author{
  John Stephan\textsuperscript{1}\thanks{Correspondence to: \texttt{john.stephan@huawei.com}},
  Razvan Puscasu\textsuperscript{1},
  Xu Yong\textsuperscript{2},
  Ding Jiutao\textsuperscript{1},
  Guan Zixuan\textsuperscript{2}\\[0.5em]
  \textsuperscript{1}Huawei, Switzerland\\
  \textsuperscript{2}Huawei, Beijing, China
}

\date{}

\begin{document}
\maketitle
 
\begin{abstract}
In expert-parallel mixture-of-experts (MoE) serving, expert computation in each MoE layer is preceded by dispatch and followed by combine, both of which lie on the critical path: tokens must be dispatched to the accelerators that host their selected experts, and the expert outputs must be combined back at the token's owner.
The cost of these collectives can dominate layer latency, particularly in multi-node deployments, where inter-node bandwidth is far below intra-node bandwidth.
We present INC-MoE, an expert-parallel MoE layer for the scale-out setting that offloads both communication phases to the network switch and overlaps them with each other and with expert computation inside a single persistent kernel. INC-MoE expresses dispatch as an in-network AllGather, realized by switch-side multicast, and combine as an in-network ReduceScatter, realized by switch-side reduction. We implement both collectives with message semantics, as message exchanges between the end hosts and the switch, and show that because the in-network AllGather and ReduceScatter load opposite directions of each link, running them concurrently reduces the bandwidth-bound communication time by $\frac{N-2}{2(N-1)}$ relative to concurrent ring-based collectives, i.e., by 33\% for four ranks and 43\% for eight.

On each end host, a single kernel partitions the tokens of a layer into bands and processes them through a three-stage AllGather–feed-forward network (FFN)–ReduceScatter pipeline.
At steady state, role-specialized thread blocks gather the next band, execute expert computation on the current band, and reduce the previous band concurrently, thereby overlapping both collectives with each other and with computation, using entirely device-side synchronization.

INC-MoE is a drop-in replacement for the MoE layer of vLLM. We evaluate it on a testbed that emulates a four-node scale-out deployment, in which a dedicated accelerator emulates an in-network computing (INC) switch. On prefill of OLMoE with four-way expert parallelism, INC-MoE reduces time to first token (TTFT) by up to 33\% and increases throughput by up to 50\% relative to vLLM's default collective path, while matching its outputs.
\end{abstract}

\section{Introduction}
\label{sec:intro}

Mixture-of-Experts (MoE) architectures~\cite{shazeer2017moe,lepikhin2021gshard,fedus2022switch} decouple a model's parameter count from its per-token compute by routing each token to a small subset of $k$ expert feed-forward networks (FFNs) out of $E$. This sparsity has made MoE widely adopted in large language models~\cite{deepseekv3,muennighoff2025olmoe}.
For large MoE models and throughput-oriented deployments, experts are often partitioned across multiple accelerators, potentially spanning multiple nodes, using \emph{expert parallelism} (EP).
Every MoE layer must then first deliver each token to the accelerators that host its selected experts (\emph{dispatch}) and afterwards bring the weighted expert outputs back to the accelerator that owns the token (\emph{combine}). Both steps sit on the critical path of every layer, and the share of time they consume grows with the degree of EP~\cite{zhang2025comet,aimuyo2025flashmoe}.

In the data-parallel attention plus expert-parallel FFN configuration used by serving engines such as vLLM~\cite{kwon2023vllm}, dispatch and combine are commonly expressed as two collectives: an \emph{AllGather} (\AG{}) that makes every rank's tokens visible to every other rank, and a \emph{ReduceScatter} (\RS{}) that sums the partial expert outputs computed on each rank and returns to every rank the rows of its own tokens~\cite{tdc2026agrs}.
In the default implementation, the MoE layer executes in three sequential stages: an \AG{} issued through the platform's collective library~\cite{hccl, nccl}, a sequence of MoE compute kernels (e.g., alignment/permutation, two grouped GEMMs with an activation in between), and an \RS{}.
In the inter-node regime, where inter-node bandwidth is much lower than intra-node bandwidth, communication can account for a substantial fraction of MoE layer latency, making the layer communication-bound.

This paper starts from two observations.
First, the \AG{} and \RS{} formulation of dispatch and combine maps naturally to operations supported by a switch capable of in-network computing (INC): AllGather can be implemented through multicast, and ReduceScatter through reduction.
INC for collectives has a long history in HPC~\cite{graham2016sharp,desensi2021flare} and distributed training~\cite{sapio2021switchml,lao2021atp}, where it has been applied mostly to all-reduce invoked as a standalone collective (\cref{sec:bg-inc}).
Recent work has also explored in-switch support for MoE communication~\cite{zhang2026dysharp}.
Second, offloading the collectives alone is not enough.
In the bandwidth-bound regime, an in-network \AG{} or \RS{} taken in isolation completes in the same time as its ring counterpart, which is already bandwidth-optimal~\cite{chan2007collective}: it still saturates one direction of each end host's link (\cref{sec:inc-cost}).
The two in-network collectives, however, saturate \emph{opposite} directions, so running them concurrently uses both directions of every link at once.
By contrast, overlapping ring \AG{} and \RS{} does not reduce their combined bandwidth-bound communication time, since each collective already saturates both directions.
Realizing this gain requires restructuring the MoE layer so that the \AG{} and the \RS{} are in flight at the same time, and so that expert computation proceeds while both are running.

We present \sys{}, an EP MoE layer for the scale-out setting built on these observations. \sys{} offloads dispatch and combine to in-network \AG{} and \RS{}, and fuses the two collectives with expert computation into one persistent kernel that pipelines all three phases at sub-layer granularity. The tokens of the layer are split into \emph{bands}; while the expert FFN runs on one band, the switch is gathering the next band and reducing the previous one.
Making this overlap effective raises three challenges that shape our design:
\begin{enumerate}[leftmargin=*,itemsep=2pt]
  \item \textbf{Fine-grained in-network collectives.}
  The \AG{} and \RS{} must be issued and completed directly from device code at band granularity, with consistent message layouts and band boundaries across ranks despite unequal per-rank token counts.

  \item \textbf{Device-side synchronization.}
  Gather, compute, and reduce thread blocks must coordinate data dependencies while progressing independently, and reductions must remain synchronized across ranks without introducing unnecessary serialization.

  \item \textbf{Pipeline balance and tail overhead.}
  The relative costs of gathering, computing, and reducing vary with the workload, requiring an appropriate choice of band size and thread-block allocation. Moreover, the loss of overlap during pipeline fill and drain must be mitigated to minimize exposed communication latency.
\end{enumerate}

We prototype the switch on a dedicated accelerator that emulates an INC-capable switch. The end-host kernel depends on the switch only through the message interface of \cref{sec:inc}, so it is unchanged by this choice.

\begin{figure}[ht!]
    \centering
    \includegraphics[width=0.7\linewidth]{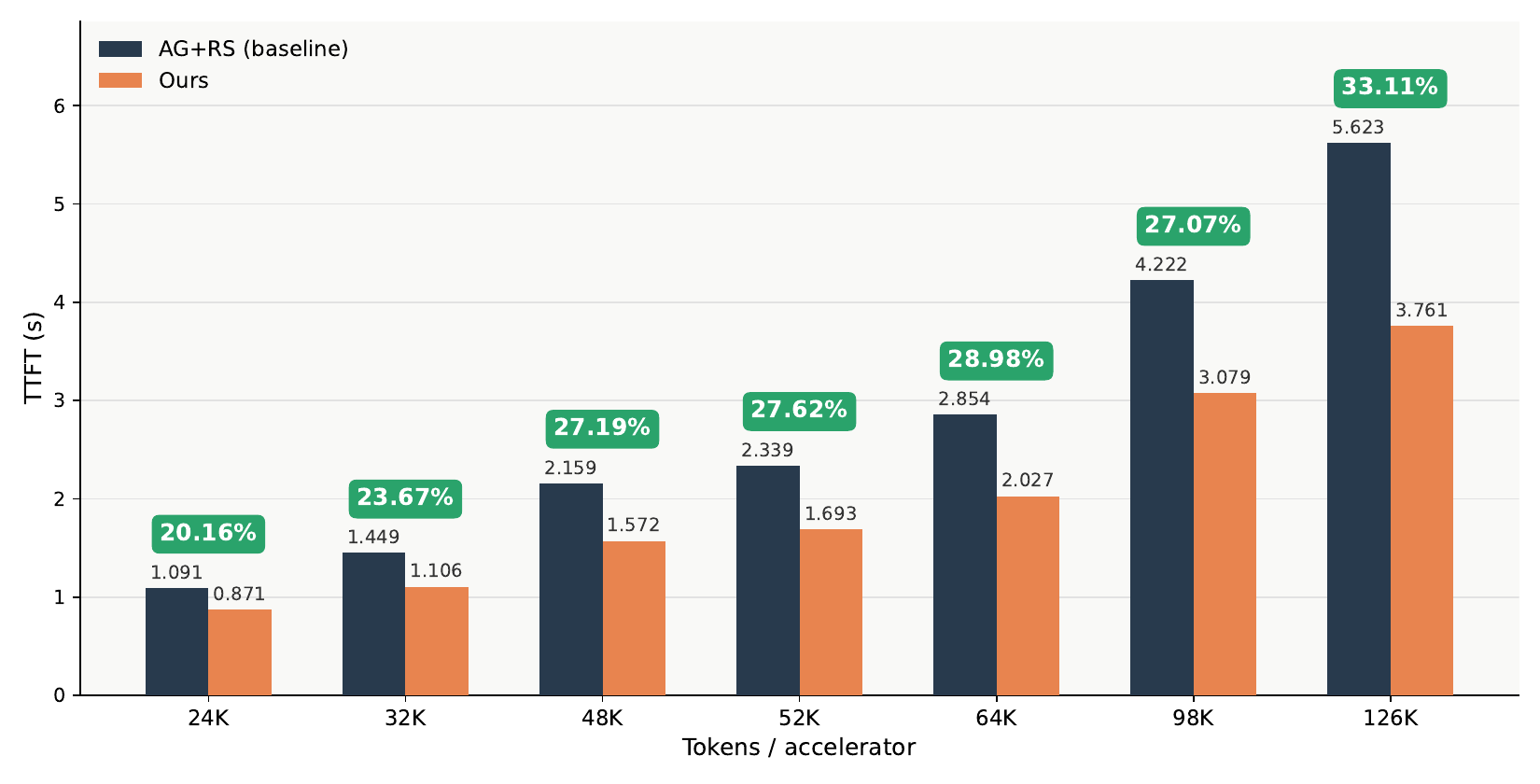}
    \caption{
        Time to first token (TTFT) of OLMoE-1B-7B prefill in vLLM with four-way data-parallel attention and four-way expert parallelism in an emulated four-node deployment, as a function of the number of prompt tokens per accelerator.
        The baseline is vLLM's default expert-parallel path, in which a ring AllGather and ReduceScatter are issued through the default collective library and surround the MoE compute kernels.
        \sys{} executes both collectives in the network and overlaps them with each other and with expert computation in a single band-pipelined kernel. Percentages give the TTFT reduction, $(t_{\mathrm{base}} - t_{\mathrm{ours}})/t_{\mathrm{base}}$, which ranges from 20.2\% at 24K tokens to 33.1\% at 126K tokens.}
    \label{fig:e2e-ttft}
\end{figure}

\paragraph{Contributions.} This paper makes the following contributions:
\begin{itemize}[leftmargin=*,itemsep=2pt]
  \item \textbf{In-network \AG{} and \RS{} with message semantics.} We design and implement in-network AllGather and ReduceScatter as message exchanges between end hosts and the switch. The \AG{} is built from rooted multicasts, and the \RS{} from rooted in-network reductions followed by a local final reduction at the root. Both sides are implemented: a client library through which accelerators issue and complete the collectives from device code, and a switch-side service that replicates and reduces message payloads. We show that overlapping the two in-network collectives reduces the bandwidth-bound communication time from $\frac{2(N-1)}{N}\cdot\frac{S}{BW}$ for ring collectives to $\frac{S}{BW}$, an improvement of $\frac{N-2}{2(N-1)}$, which is 33.3\% for $N{=}4$ and 42.9\% for $N{=}8$ (\cref{sec:inc,sec:inc-cost}).
  \item \textbf{A fused kernel that makes the two collectives overlap.} The bound above is attained only when an \AG{} and an \RS{} are in flight at the same time, but within an MoE layer they are strictly dependent: the \RS{} needs the FFN output, which in turn needs the \AG{}. We break this dependency chain by splitting the layer's tokens into bands and fusing \AG{}, expert FFN, and \RS{} into one persistent kernel whose thread blocks are specialized into gather, compute, and reduce roles. At steady state, the \AG{} of one band, the FFN of a second, and the \RS{} of a third can run concurrently, so the two collectives overlap with each other and with expert computation. The FFN writes its router-weighted output directly into the \RS{} send buffer, and the reduced output is handed to the serving engine in place, so no copy separates computation from communication. Because the overlap vanishes once the last \AG{} completes, gather blocks then join the reduction of the final bands (\emph{tail assistance}), shortening the part of the layer where the \RS{} runs alone.
  \item \textbf{Integration and evaluation.} \sys{} is a drop-in replacement for vLLM's MoE layer, requiring no changes to model definitions (\cref{sec:integration}).
  On OLMoE prefill with four-way EP, in a testbed that emulates four nodes and the INC switch (\cref{sec:eval-setup}), it reduces TTFT by up to 33.1\% and increases prefill throughput by up to 49.5\% relative to vLLM's default path (\cref{fig:e2e-ttft,fig:e2e-tps}), with per-token cosine similarity of $1.0$ to the baseline across all layers and ranks.
  In a standalone benchmark with concurrent AllGather and ReduceScatter, the in-network collectives achieve 25.6--31.8\% higher equivalent bandwidth than concurrently executed ring collectives (\cref{sec:eval-bw}).
\end{itemize}


\section{Background}
\label{sec:background}

\subsection{MoE Layers and Expert Parallelism}
\label{sec:bg-moe}

An MoE layer replaces the dense FFN of a transformer block with $E$ expert FFNs and a router. For a token with hidden state $x_t \in \mathbb{R}^{H}$, the router selects a set $\mathcal{K}(t)$ of $k$ experts with weights $g_{t,e}$, and the layer computes
\begin{equation}
  y_t = \sum_{e \in \mathcal{K}(t)} g_{t,e}\, \mathrm{FFN}_e(x_t), \qquad
  \mathrm{FFN}_e(x) = W^{\mathrm{down}}_e \big( \mathrm{SiLU}(W^{\mathrm{gate}}_e x) \odot W^{\mathrm{up}}_e x \big),
  \label{eq:moe}
\end{equation}
with $W^{\mathrm{gate}}_e, W^{\mathrm{up}}_e \in \mathbb{R}^{I \times H}$ and $W^{\mathrm{down}}_e \in \mathbb{R}^{H \times I}$. Under EP of degree $N$, rank $r$ hosts a subset $\mathcal{E}_r$ of $E/N$ experts. With data-parallel attention, rank $r$ also owns a batch of $T_r$ tokens, and different ranks may own different numbers of tokens.

\subsection{Dispatch and Combine as AllGather and ReduceScatter}
\label{sec:bg-agrs}

Dispatch and combine are commonly implemented with all-to-all exchanges~\cite{lepikhin2021gshard,deepep2025}.
An alternative, used by vLLM's default EP path and described in~\cite{tdc2026agrs}, expresses them with two regular collectives. The \AG{} gives every rank the hidden states of all $\sum_r T_r$ tokens. Each rank $r$ then computes, for every token, the partial output
\begin{equation}
  z^{(r)}_t = \sum_{e \in \mathcal{K}(t) \cap \mathcal{E}_r} g_{t,e}\, \mathrm{FFN}_e(x_t),
  \label{eq:partial}
\end{equation}
which is zero if none of the token's experts is local. The \RS{} sums $z^{(r)}_t$ over $r$, which yields $y_t$ by \cref{eq:moe}, and delivers to each rank the rows of the tokens it owns.

\paragraph{Payload efficiency.} The \AG{} sends each token to every other rank regardless of routing, whereas an ideal all-to-all that deduplicates per destination sends it only to ranks that host at least one of its experts. For top-$k$ routing with $k$ large relative to $N$, the two approaches differ little. Under uniform routing, the probability that a given rank hosts none of a token's $k$ experts is $\binom{E - E/N}{k} / \binom{E}{k}$. For $E{=}64$, $k{=}8$ and $N{=}4$, this probability is $8.5\%$, so a token is needed by $2.74$ of the $3$ remote ranks in expectation, and the \AG{} sends only about $9\%$ more data than a fully deduplicated all-to-all. The same argument applies to the \RS{}. What the \AG{}/\RS{} formulation gives up in payload, it repays with a regular and routing-independent communication pattern, which is what makes it amenable to switch offload.

\subsection{In-network Collectives}
\label{sec:bg-inc}

In-network computing (INC) executes collective operations in the switches of the fabric rather than on the end hosts~\cite{graham2016sharp,sapio2021switchml,lao2021atp}. A switch that can replicate an incoming message to several output ports implements a multicast, and a switch that can combine the payloads of several incoming messages implements a reduction. Compared with collectives executed by the end hosts, in-network collectives reduce the number of times data crosses the end-host links, replace the dependent steps of ring and tree algorithms with a single pass through the switch, and move the reduction arithmetic off the end hosts.
Prior work has applied INC mainly to all-reduce, both in HPC~\cite{graham2016sharp,desensi2021flare} and in distributed training~\cite{sapio2021switchml,lao2021atp,gebara2021panama,liu2023netreduce,li2019iswitch}. Only recently has INC been considered for the dispatch and combine of expert parallelism~\cite{zhao2025deepseekhw,zhang2026dysharp}, and, to our knowledge, it has not yet been applied to expert-parallel prefill across nodes within a serving engine, the setting \sys{} targets.
\Cref{sec:inc} describes how \sys{} builds \AG{} and \RS{} from multicast and reduction, and \cref{sec:inc-cost} analyzes their cost for expert parallelism.

\subsection{Device-initiated Communication}
\label{sec:bg-device}
In the host-side APIs of conventional collective libraries~\cite{mpi2021,nccl,rccl,hccl}, collectives are launched from the host as separate operations, which forces the host to sequence communication and computation. Overlap is then limited to the granularity at which the host can split the work into separate launches, and every additional split adds launch overhead and cross-stream synchronization. Fine-grained pipelining therefore pays a growing fixed cost per stage: a host-driven version of the band pipeline in \cref{sec:pipeline} would require, for each MoE layer, one \AG{} and one \RS{} launch per band, interleaved with the expert kernels and ordered by events. Device-initiated communication~\cite{mscclpp2026, hamidouche2025gpuinitiatednetworkingnccl} removes the host from this loop. A running kernel can post transfers, signal peers, and wait on completions itself, so communication starts as soon as the data it depends on has been produced, at whatever granularity the kernel chooses, and without returning control to the host~\cite{deepep2025,aimuyo2025flashmoe}. This makes it possible to place communication and computation in the same persistent kernel. \sys{} relies on this capability to issue the in-network collectives of each band of tokens from the device, overlapping them with each other and with expert computation inside a single kernel launch per MoE layer.


\section{Design of \sys{}}
\label{sec:design}

\begin{figure}[ht!]
    \centering
    \includegraphics[width=0.6\linewidth]{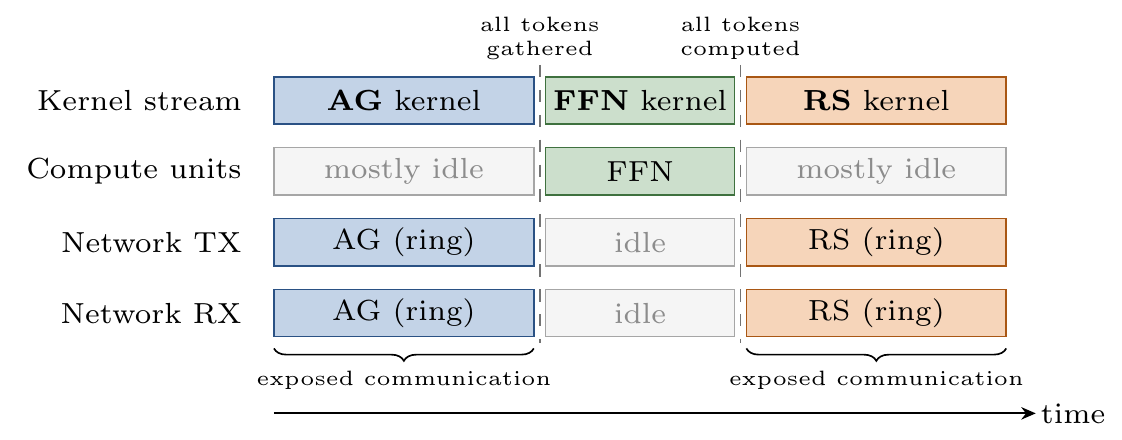}
    \caption{Execution of an MoE layer under the default EP path.}
    \label{fig:default-path}
\end{figure}

\paragraph{Motivation.}
\Cref{fig:default-path} illustrates how an MoE layer executes under the default
vLLM EP path.
Two inefficiencies follow from this execution. First, the three stages are serialized: the FFN cannot start until the whole \AG{} has completed, and the \RS{} cannot start until the FFN has finished on every token, so the full communication time is exposed on the critical path.
Second, the collectives run on the end hosts as ring algorithms, which load both directions of every link equally. Each collective therefore saturates the link on its own, and overlapping the \AG{} with the \RS{} does not shorten their combined time (see \cref{sec:inc-cost} and \cref{sec:eval-bw}).

Neither remedy suffices on its own.
Offloading the collectives to the switch, by itself, leaves them serialized with each other and with the FFN. Overlapping endpoint collectives with computation, as in micro-batch overlap~\cite{vllmdbo,zheng2024sglang} or fused kernels~\cite{zhang2025comet,aimuyo2025flashmoe}, hides communication behind computation but does not shorten the communication itself. \sys{} combines the two: the switch executes the collectives, which makes them overlappable with each other (\cref{sec:inc-cost}), and the accelerator overlaps them with each other and with computation at the granularity of bands.

\begin{figure}[ht!]
    \centering
    \includegraphics[width=0.7\linewidth]{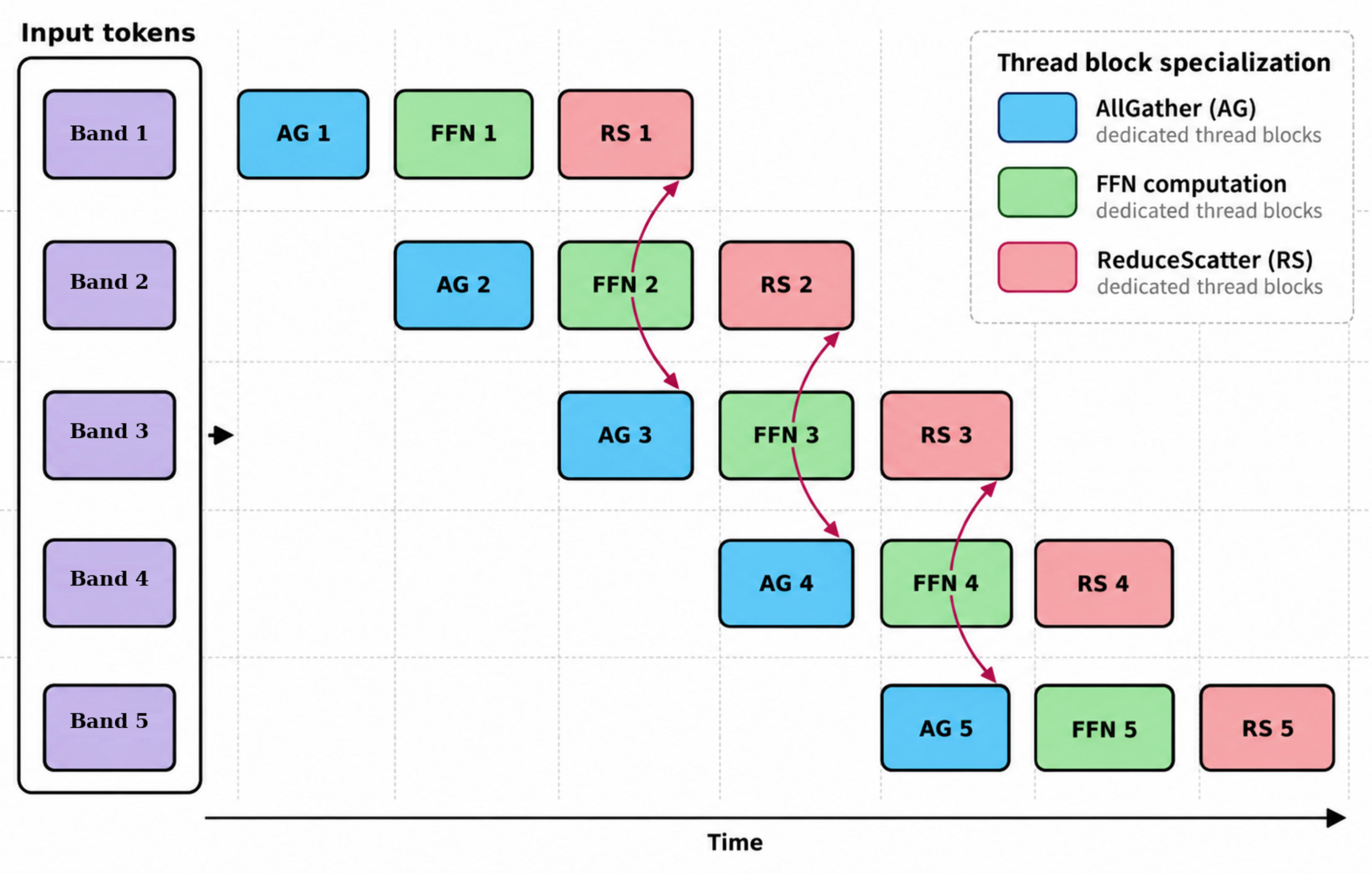}
    \caption{
        Overview of \sys{}: pipelined communication--computation overlap (AG/FFN and RS/FFN) and communication--communication overlap (AG/RS) across bands.
        Dedicated thread blocks are specialized for AG, FFN computation, and RS, enabling concurrent execution of communication and computation tasks.
    }
    \label{fig:overview}
\end{figure}

\paragraph{Overview.}
\Cref{fig:overview} gives an overview of \sys{}. The EP ranks reside on different nodes and communicate over the scale-out network through a switch.
The layer receives the hidden states of the rank's tokens and the router output (selected experts and weights) from the serving engine and returns the combined output for the same tokens. Internally, it launches one persistent kernel per MoE layer. The kernel's thread blocks are divided into three roles: \emph{gather} blocks issue in-network \AG{}s, \emph{compute} blocks run the expert FFNs, and \emph{reduce} blocks issue in-network \RS{}s and complete them with a local reduction. The roles process the tokens band by band and hand data to each other through device memory.

\subsection{In-Network AllGather and ReduceScatter with Message Semantics}
\label{sec:inc}
\begin{figure*}[ht!]
    \centering
    \includegraphics[width=\textwidth]{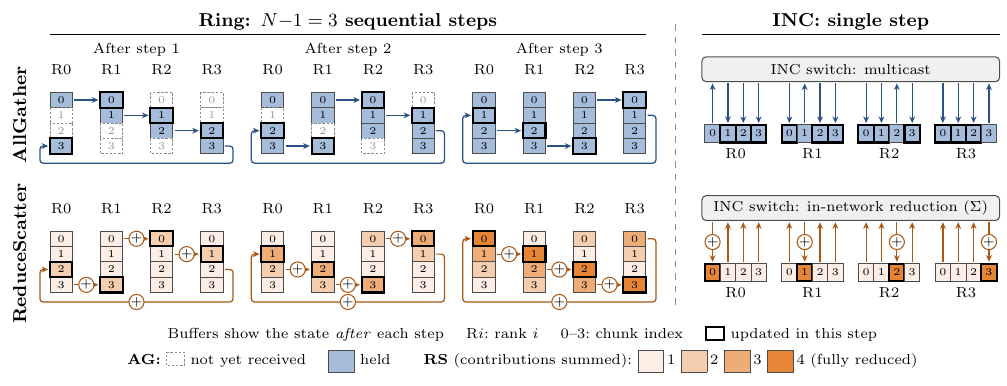}
    \caption{Ring vs.\ in-network (\sys{}) \AG{} (top) and \RS{} (bottom) for
    $N{=}4$ ranks. Chunk $r$ is rank $r$'s block of the gather/reduce buffer.
    Buffers show the state after each step. The ring needs $N{-}1$ sequential
    steps, with a reduction on the receiving accelerator ($\oplus$) in every
    \RS{} step. \sys{} completes each collective in a single step: the switch
    multicasts each root's chunk for the \AG{}, and for the \RS{} it reduces the
    $N{-}1$ non-root contributions in the network ($\Sigma$), leaving the root
    a single local reduction with its own chunk.}
    \label{fig:ring-inc}
\end{figure*}

\sys{} implements the \AG{} and \RS{} with \emph{message semantics}. End hosts never read or write each other's memory. Every transfer is a message exchanged between an end host and the switch, and the switch is responsible for replicating and reducing message payloads. This contrasts with memory-semantic designs, in which each end host writes directly into the memory of its peers~\cite{aimuyo2025flashmoe,deepep2025}.
Combining message semantics with in-network computing enables the switches to deliver one incoming message to several destinations or to combine several incoming messages into one, in the scale-out setting.
\Cref{fig:ring-inc} contrasts the resulting collectives with the ring algorithms used by the default path.

\paragraph{End-host interface.} The end-host side of the collectives is a client library callable from device code. A thread block posts an \AG{} or \RS{} on a region of a registered buffer, continues with other work, and later waits on the completion of that request.
Posting a request does not move data on the accelerator: the network interface reads the payload from the registered buffer, transmits it to the switch, and writes incoming payloads directly into registered memory.
In our deployment, messages are carried over RoCEv2~\cite{ibta2014rocev2,guo2016rdma}.
The accelerator's thread blocks are thus only involved where data must be prepared before transmission or processed after reception.
The library also provides a cross-rank barrier. Each message carries, besides its payload, information that the switch and the receiving end host need to process it (e.g., the operation, the root, the byte range of the payload within the registered buffer, the address at which the receiver must signal completion).
The end-host code depends on the switch only through this interface, so the same kernel can target any switch that implements the \AG{} and \RS{} message semantics.

\paragraph{AllGather.} The \AG{} is built from rooted multicasts, one per rank. Each rank acts as the root for its own contribution:
\begin{enumerate}[leftmargin=*,itemsep=1pt]
  \item The root sends its contribution to the switch as a single message.
  \item The switch multicasts the message to the remaining $N{-}1$ ranks.
   \item The network interface of each receiving rank writes the payload at the byte range given in the message and signals the completion address.
\end{enumerate}
All $N$ ranks act as roots concurrently. A rank's \AG{} is complete once the contributions of the $N{-}1$ remote roots have arrived. Its own contribution requires no network transfer.
Each rank thus sends its contribution once, regardless of $N$, while a ring \AG{} would forward the contributions of $N{-}1$ ranks over $N{-}1$ sequential steps (\cref{fig:ring-inc}, top left).

\paragraph{ReduceScatter.} The \RS{} is built from rooted reductions, one per output shard. Each rank is the root of the shard that holds its own tokens:
\begin{enumerate}[leftmargin=*,itemsep=1pt]
  \item The $N{-}1$ non-root ranks send their partial results for the root's shard to the switch, chunk by chunk.
  \item The switch reduces each chunk as soon as all $N{-}1$ non-root ranks have delivered it, reducing different chunks in parallel.
  \item The switch forwards each reduced chunk to the root as soon as it is ready, so arrival, reduction, and forwarding are pipelined.
  \item The root adds its own partial result to the aggregated result locally, obtaining the final shard, and signals completion.
\end{enumerate}

Each rank therefore participates in $N$ rooted reductions: as a sender in the $N{-}1$ reductions rooted at the other ranks, and as the root of its own.
Because the root keeps its own contribution local, a rank sends $N{-}1$ shards instead of all $N$, and the switch reduces $N{-}1$ contributions per shard instead of $N$.
A rank receives a single already reduced shard, whereas a ring \RS{} would receive and reduce $N{-}1$ partial shards on the end host, one per step (\cref{fig:ring-inc}, bottom left).

\paragraph{Data layout.} Ranks own different numbers of tokens, but the byte ranges carried by messages must refer to the same elements on every rank. \sys{} therefore lays out the gather and reduce buffers (on every rank) as $N$ per-rank blocks of $T_{\max} = \max_r T_r$ rows each, where rank $r$'s tokens occupy the first $T_r$ rows of block $r$. The token counts of all ranks are known before the layer runs (vLLM tracks them for data-parallel execution), so every rank derives the same layout without exchanging metadata.

\paragraph{Concurrent operation.} The band pipeline (\cref{sec:pipeline}) keeps several collectives in flight at once, for example the \AG{} of one band and the \RS{} of an earlier band, and issues collectives of the same kind back to back. Completion signals are cumulative and compared against monotonically increasing per-operation thresholds (epochs) rather than being reset between operations, which removes the race between a reset and a late signal from the previous operation.

\subsection{Communication Cost}
\label{sec:inc-cost}
We now compare the in-network collectives with ring collectives executed by the end hosts in terms of communication cost.
Let $S$ denote the size in bytes of the full gathered buffer, so that each of the $N$ ranks contributes $S/N$ bytes to the \AG{} and owns $S/N$ bytes of the \RS{} output. We assume full-duplex links with bandwidth $BW$ per direction between each end host and the network, a switch that forwards and reduces at line rate, and messages large enough for the bandwidth term to dominate. \Cref{tab:traffic} lists the bytes each rank sends and receives.

\begin{table}[ht!]
  \centering
  \caption{Per-rank traffic and bandwidth-bound time of AllGather and ReduceScatter for $N$ ranks, in units of $S$ (traffic) and $S/BW$ (time), where $S$ is the size of the full gathered buffer. The time of each configuration is set by the more heavily loaded link direction.}
  \label{tab:traffic}
  \small
  \begin{tabular}{lcccccc}
    \toprule
    & \multicolumn{2}{c}{AllGather} & \multicolumn{2}{c}{ReduceScatter} & \multicolumn{2}{c}{AG + RS (overlap)} \\
    \cmidrule(lr){2-3} \cmidrule(lr){4-5} \cmidrule(lr){6-7}
    & Ring & In-network & Ring & In-network & Ring & In-network \\
    \midrule
    Sent per rank     & $\frac{N-1}{N}$ & $\frac{1}{N}$   & $\frac{N-1}{N}$ & $\frac{N-1}{N}$ & $\frac{2(N-1)}{N}$ & $1$ \\[2pt]
    Received per rank & $\frac{N-1}{N}$ & $\frac{N-1}{N}$ & $\frac{N-1}{N}$ & $\frac{1}{N}$   & $\frac{2(N-1)}{N}$ & $1$ \\[2pt]
    Time              & $\frac{N-1}{N}$ & $\frac{N-1}{N}$ & $\frac{N-1}{N}$ & $\frac{N-1}{N}$ & $\frac{2(N-1)}{N}$ & $1$ \\
    \bottomrule
  \end{tabular}
\end{table}

\paragraph{Each collective alone.} A ring \AG{} or \RS{} sends and receives $\frac{N-1}{N}S$ bytes per rank and completes in $\frac{N-1}{N}\cdot\frac{S}{BW}$. The in-network \AG{} reduces the bytes a rank sends to $\frac{S}{N}$, but every rank must still receive the $\frac{N-1}{N}S$ bytes of the other ranks. Symmetrically, the in-network \RS{} reduces the bytes a rank receives to $\frac{S}{N}$, but every rank must still send its $\frac{N-1}{N}S$ bytes of partial results. In the bandwidth-bound regime, each in-network collective taken alone therefore completes in the same time as its ring counterpart.
Its benefits are the removal of the $N{-}1$ dependent ring steps and the reduction of the end-host arithmetic from $N{-}1$ partial shards to a single local addition at the root, which matter mainly for small messages and for the compute resources left to the FFN.

\paragraph{Overlapping the two collectives.} The in-network \AG{} is receive-heavy and the in-network \RS{} is send-heavy; this asymmetry has also been observed by~\cite{zhang2026dysharp}. When the two run concurrently, each direction of a rank's link carries exactly
\begin{equation}
  \frac{S}{N} + \frac{N-1}{N}\,S \;=\; S
\end{equation}
bytes, so the pair completes in $S/BW$. The ring collectives, by contrast, already load both directions equally, so overlapping a ring \AG{} with a ring \RS{} leaves $\frac{2(N-1)}{N}S$ bytes in each direction and does not shorten their combined time. The communication latency of the two configurations is
\begin{equation}
  t_{\mathrm{ring}} = \frac{2(N-1)}{N}\cdot\frac{S}{BW}, \qquad
  t_{\mathrm{INC}} = \frac{S}{BW},
  \label{eq:latency}
\end{equation}
and the theoretical improvement of the overlapped in-network collectives is
\begin{equation}
  1 - \frac{t_{\mathrm{INC}}}{t_{\mathrm{ring}}} \;=\; \frac{N-2}{2(N-1)}.
  \label{eq:improvement}
\end{equation}
The improvement is zero for $N{=}2$, where each in-network collective already uses its link symmetrically. It reaches 33.3\% for $N{=}4$ and 42.9\% for $N{=}8$, and approaches 50\% as $N$ grows. Equivalently, the overlapped in-network collectives can deliver up to $\frac{2(N-1)}{N}$ times the equivalent bandwidth of concurrent ring collectives, i.e., $1.5\times$ for $N{=}4$. Attaining this bound requires keeping an \AG{} and an \RS{} in flight simultaneously for as much of the layer as possible, which is the purpose of the kernel described next.

\subsection{Overlap Kernel: Band Pipeline with Role-Specialized Thread Blocks}
\label{sec:pipeline}

\paragraph{Bands.} \sys{} splits each rank's tokens into $n_b$ bands of $\beta = \lceil T_{\max} / n_b \rceil$ rows. Band $j$ consists of rows $[j\beta, \min((j{+}1)\beta, T_{\max}))$ of every rank's block, so a band covers a slice of every rank's tokens and is self-contained for all three phases. The \AG{} of band $j$ gathers the $j$-th slice of every rank, the FFN of band $j$ computes the partial outputs (\cref{eq:partial}) of exactly those tokens, and the \RS{} of band $j$ reduces them and returns to each rank its own $j$-th slice. All three roles and all ranks compute band boundaries with the same arithmetic. This invariant is essential for correctness: a single off-by-one difference between roles leads the reduce blocks to send rows the compute blocks have not written, or the compute blocks to read rows the gather blocks have not received.

\paragraph{Role specialization.} The kernel is launched with $B_{\AG} + B_{\mathrm{FFN}} + B_{\RS}$ blocks, and each block determines its role from its index. Although the network interface performs the transfers, both communication roles do work on the accelerator. Gather blocks iterate over bands, copy the rank's tokens of each band into the registered send buffer at the offsets of the padded layout (\cref{sec:inc}), post the \AG{} of the band, and publish its completion. This work is light, so few gather blocks suffice. Compute blocks iterate over bands, wait for the corresponding \AG{}, and run the expert FFNs of the band (\cref{sec:compute}).
Reduce blocks iterate over bands, wait for the partial outputs of the band, post its \RS{}, and, once the aggregated result arrives from the switch, perform the root's local reduction: they add the partial sum received from the switch to the rank's own partial results, in place in the rank's block of the \RS{} buffer $Z$, which then holds the final output.
This local reduction is a memory-bound computation over the rank's entire output, and it is the main reason the reduce role needs dedicated blocks. Each role advances independently at its own pace, as shown in \cref{alg:pipeline}.

\paragraph{Communication--communication and communication--computation overlap.} At steady state, three bands are in flight at once: the gather blocks run the \AG{} of band $j{+}1$, the compute blocks run the FFN of band $j$, and the reduce blocks run the \RS{} of band $j{-}1$. The pipeline therefore realizes two kinds of overlap. \emph{Communication--communication} overlap keeps an in-network \AG{} and an in-network \RS{} in flight at the same time, which is precisely the schedule that \cref{sec:inc-cost} shows to reduce the communication time from $\frac{2(N-1)}{N}\cdot\frac{S}{BW}$ to $\frac{S}{BW}$: the \AG{} keeps the receive direction of each link busy while the \RS{} keeps the send direction busy. \emph{Communication--computation} overlap hides the expert FFN behind the two collectives. Because the gather and reduce roles are assigned to separate groups of blocks that progress independently, the two collectives overlap without any coordination beyond the per-band synchronization described below.

\begin{algorithm}[t]
  \caption{Band pipeline executed by one rank. Each role runs in its own set of thread blocks.}
  \label{alg:pipeline}
  \begin{algorithmic}[1]
    \Statex \textbf{Gather blocks}
    \For{$j = 0, \dots, n_b - 1$}
      \State copy the rank's tokens of band $j$ into the registered send buffer
      \State post in-network \AG{} of band $j$ into the gather buffer $\hat{X}$; wait for completion
      \State \textsc{Release}($\mathit{agReady}[j]$)
    \EndFor
    \If{tail assistance is enabled} \State join the reduce blocks for bands $j \ge n_b - n_{\mathrm{tail}}$ \EndIf
    \Statex \textbf{Compute blocks}
    \For{$j = 0, \dots, n_b - 1$}
      \State \textsc{Acquire}($\mathit{agReady}[j]$)
      \For{each local expert $e \in \mathcal{E}_r$}
        \State $R \gets$ rows of band $j$ routed to $e$; \textbf{if} $R = \emptyset$ \textbf{then continue}
        \State $A \gets \mathrm{SwiGLU}\big(\hat{X}[R]\, [W^{\mathrm{gate}}_e; W^{\mathrm{up}}_e]^\top\big)$ \Comment{activation fused in epilogue}
        \State $Z[R] \mathrel{+}= g_{R,e} \odot \big(A\, {W^{\mathrm{down}}_e}^\top\big)$ \Comment{weighted store into the \RS{} send buffer}
      \EndFor
      \State advance this block's progress counter to $j+1$ \Comment{release}
    \EndFor
    \Statex \textbf{Reduce blocks}
    \For{$j = 0, \dots, n_b - 1$}
      \State wait until all compute progress counters reach $j+1$ \Comment{acquire}
      \State cross-rank band barrier
      \State post in-network \RS{} of band $j$ from $Z$; wait for the partial sum from the switch
      \State $Z[\text{own rows of band } j] \mathrel{+}= \text{partial sum}$ \Comment{root's local reduction, in place}
    \EndFor
  \end{algorithmic}
\end{algorithm}

\paragraph{Tail assistance.} The overlap between \AG{} and \RS{} disappears at both ends of the layer. The \AG{} of the first band runs before any \RS{} can start, and the \RS{}s of the last bands run after the final \AG{} has completed. In this drain phase, the gather blocks sit idle, while the \RS{}s of the last bands, including their local reductions, lie directly on the critical path.
With tail assistance, gather blocks that have posted their last \AG{} join the reduce blocks for the final $n_{\mathrm{tail}}$ bands and take over part of their local reductions, so these bands are processed by $B_{\AG} + B_{\RS}$ blocks instead of $B_{\RS}$.
Tail assistance is most effective at large token counts (64K tokens per accelerator and above). Below that, its effect is small.

\paragraph{Synchronization within an accelerator.} Roles hand data to each other through flags in device memory written with release semantics and read with acquire semantics, so that a consumer observes all data written before the flag. Rather than gating each step on a barrier across all blocks, every block that produces work maintains its own monotonically increasing progress counter, and a consumer waits only on the counters of the blocks whose output it needs. In the reduce role, this lets each reduce block advance through its share of a band at its own pace, instead of preceding every reduction chunk by a barrier across the whole grid (which makes the slowest reduce block set the pace of all of them). Because counters only increase, they never need to be reset within a launch, which avoids reset races of the kind described in \cref{sec:inc}.

\paragraph{Synchronization across accelerators.} The \RS{} of band $j$ combines the partial outputs of all ranks, so reduce blocks on different ranks must not drift apart by more than one band. \sys{} enforces this with a cross-rank barrier at band granularity, issued before each band's \RS{}. Only a single block per rank participates in the cross-rank barrier and broadcasts its completion to the other blocks of the rank through a local flag.

\paragraph{Empty bands.} When ranks own different numbers of tokens, a band can be empty on some ranks, and a band can contain no rows routed to a given local expert. Blocks skip the work in these cases, but they must still advance every counter and participate in every barrier exactly as on the non-empty path. Otherwise, the number of barrier arrivals differs across blocks or ranks and the pipeline deadlocks.

\subsection{Pipeline Configuration}
\label{sec:balance}

The throughput of the band pipeline is set by its slowest stage, and which stage is slowest depends on the number of tokens per accelerator. More bands shorten the fill and drain phases, during which \AG{} and \RS{} do not overlap, but add per-band fixed costs: collective latency, synchronization, and smaller GEMMs.

The number of bands $n_b$, the allocation of blocks to roles $(B_{\AG}, B_{\mathrm{FFN}}, B_{\RS})$, and the use of tail assistance were determined through an exhaustive empirical study, which swept these parameters across a range of per-accelerator token counts. The study showed that the optimal configuration shifts with the number of tokens per accelerator, as the balance between communication and compute changes. \sys{} therefore applies the best-performing configuration for each token-count regime at runtime. The number of compute blocks is a compile-time parameter of the kernel, so each regime uses a kernel variant specialized for its compute allocation.

\subsection{Fused Expert Computation}
\label{sec:compute}
For each band and each local expert, the compute blocks select the rows of the band routed to the expert and run two GEMMs: the gate/up projection with SwiGLU, and the down projection with router weighting. Experts are processed one after another within a band, and the compute blocks cooperate on the tiles of each GEMM.

\paragraph{GEMM.} Each (band, expert) pair is processed with two GEMMs implemented as device functions inside the persistent kernel, and local experts are processed one after another within a band rather than with a grouped GEMM. Each compute block is split into one producer group, which issues asynchronous bulk copies of operand tiles into a three-stage ring of shared-memory buffers, and two consumer groups, which issue asynchronous matrix-multiply instructions on filled stages and release them to the producer. Because the GEMM uses only block-local synchronization, it can be invoked by a subset of the kernel's blocks while the remaining blocks run communication.
Our implementation prioritizes communication efficiency over individually optimized GEMM throughput, reflecting the communication-dominated regime targeted by this work.
Grouped GEMMs over the local experts of a band are a natural extension and would increase the margin by which communication dominates.

\paragraph{Epilogue fusion.} The gate and up projections are computed by one GEMM over the concatenated weight matrix, and SwiGLU is applied to the accumulators in registers in the epilogue, so that only the activated intermediate is written to device memory. The epilogue of the down projection multiplies each row by its router weight $g_{t,e}$ and accumulates the result directly into the token's row of the \RS{} send buffer $Z$. No separate weighting, scatter or staging step is needed between the FFN and the \RS{}.
Note that the \RS{} buffer is zeroed at the start of each MoE layer, so that the contributions of a rank's local experts accumulate correctly.

\paragraph{Zero-copy output.} The root's local reduction leaves the final output for the rank's tokens in place, in the rank's own block of the \RS{} buffer $Z$. Instead of copying these rows into a separately allocated output tensor, \sys{} exposes that block to the serving engine as a tensor view, so the output of the MoE layer is consumed in place. The buffer is not reused until the engine has consumed the output.
\section{Implementation and Integration with the Serving Engine}
\label{sec:impl}
\label{sec:integration}

\paragraph{End host.} The end-host side of \sys{} consists of the fused kernel, the INC client library, and a launcher exposed to Python. Device-initiated communication is built on MSCCL++~\cite{mscclpp2026}.

\paragraph{Switch.} The switch-side service implements the message semantics of \cref{sec:inc}. In our prototype, it runs on a dedicated accelerator attached to the scale-out network, which emulates an INC-capable switch.

\paragraph{Serving-engine integration.} \sys{} replaces the MoE layer of vLLM at the boundary between the router and the expert computation. Routing (expert selection and weights) is delegated to the engine's own routing function, so routing semantics, including normalization of the top-$k$ weights, are unchanged. The per-rank token counts needed for the padded layout are taken from the engine's data-parallel metadata. Model dimensions, the number of experts, $k$, the data type, and the EP degree are runtime parameters. As a result, any MoE model whose experts are SwiGLU FFNs in fp16 or bf16 can use the layer without changes to the kernel or the model definition.

\section{Evaluation}
\label{sec:eval}

\subsection{Setup}
\label{sec:eval-setup}

\begin{figure}[ht!]
  \centering
  \includegraphics[width=0.75\linewidth]{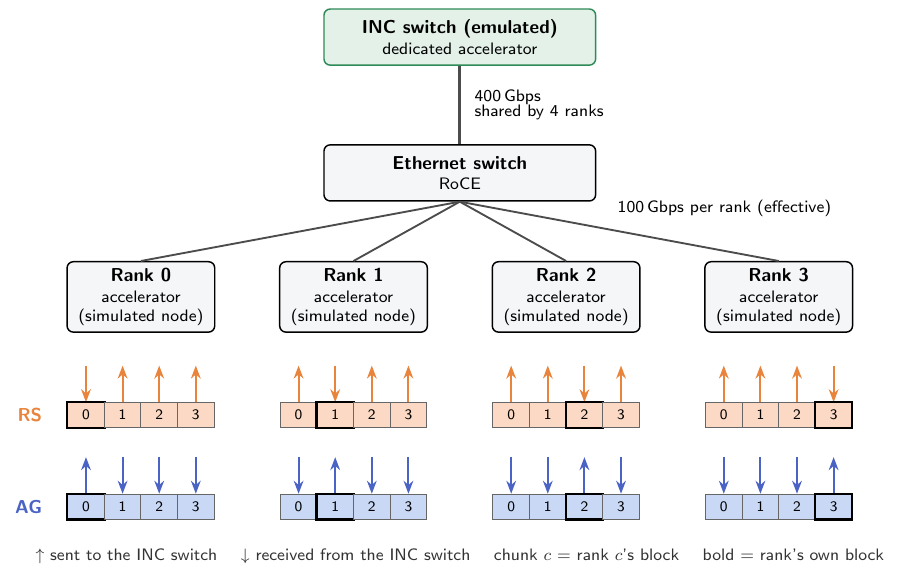}
  \caption{Testbed. Four accelerators act as EP ranks, each emulating a node, and communicate over a RoCE network through an Ethernet switch. A fifth accelerator emulates the INC switch and is attached through a 400\,Gbps link shared by the four ranks, giving each rank an effective 100\,Gbps.}
  \label{fig:testbed}
\end{figure}

\paragraph{Testbed.} We emulate a four-node scale-out deployment, shown in \cref{fig:testbed}.
The testbed uses five high-performance accelerators: four serving as expert-parallel ranks, each emulating a separate node, and a fifth emulating the INC switch.
Each accelerator supports high-throughput matrix operations and high-bandwidth device memory.
Neither system uses intra-server interconnects: all traffic of \sys{} and of the baseline crosses the accelerators' NICs and an Ethernet switch over RoCE. The emulator is attached to the Ethernet switch through a 400\,Gbps link shared by the four EP ranks, so each rank communicates with the INC switch at an effective 100\,Gbps.

\paragraph{Model and workload.} We serve OLMoE-1B-7B~\cite{muennighoff2025olmoe}, which has 16 MoE layers with $E{=}64$ experts, top-$k{=}8$ routing, hidden size $H{=}2048$, and expert intermediate size $I{=}1024$, in fp16. We run vLLM with data-parallel attention (DP${}=4$), EP${}=4$, and no tensor parallelism. We evaluate prefill and vary the aggregate number of tokens across prompts per accelerator from 24K to 126K. Each configuration is run 10 times as warm-up, followed by 30 measured repetitions, and we report the mean.
We compare the output of every MoE layer of \sys{} against the baseline on identical inputs. We report a per-token cosine similarity of $1.0$ for all 16 MoE layers on all four ranks.

\paragraph{Baseline.} The baseline is vLLM's AllGather/ReduceScatter expert-parallel path: a ring AllGather and a ring ReduceScatter, issued through the platform's default collective communication library, with vLLM's MoE compute kernels in between. Both systems run the same model weights, routing function, prompts, and DP${}=4$/EP${}=4$ configuration, and differ only in the MoE layer. The library's intra-server transports are disabled, so its collectives use its RDMA transport over the same RoCE network as \sys{}. They do not traverse the emulator's shared link. To compare both systems at the same per-rank bandwidth, we limit each baseline rank to 100\,Gbps per direction via NIC rate limiting, matching the effective bandwidth available to each rank in \sys{}.

\paragraph{Metrics.} We report time to first token (TTFT), prefill throughput per accelerator (prompt tokens divided by TTFT), and the equivalent bandwidth of the \AG{} and \RS{} collectives.

\subsection{In-Network Collective Bandwidth}
\label{sec:eval-bw}

We first evaluate the collectives in isolation. For each message size, we launch an \AG{} and an \RS{} concurrently, both for the ring baseline and for the in-network collectives of \sys{}, and measure the time until both complete.
The \emph{equivalent bandwidth} is thus the combined size of the AG and RS buffers divided by this time. In the bandwidth-bound regime it is independent of the message size and, by \cref{eq:latency}, bounded by $\frac{N}{N-1}\,BW$ for the ring collectives, which gain nothing from running concurrently, and by $2\,BW$ for the overlapped in-network collectives. With $N{=}4$ and $BW{=}100$\,Gbps per rank, these bounds are 133 and 200\,Gbps.

\begin{figure}[ht!]
    \centering
    \includegraphics[width=0.7\linewidth]{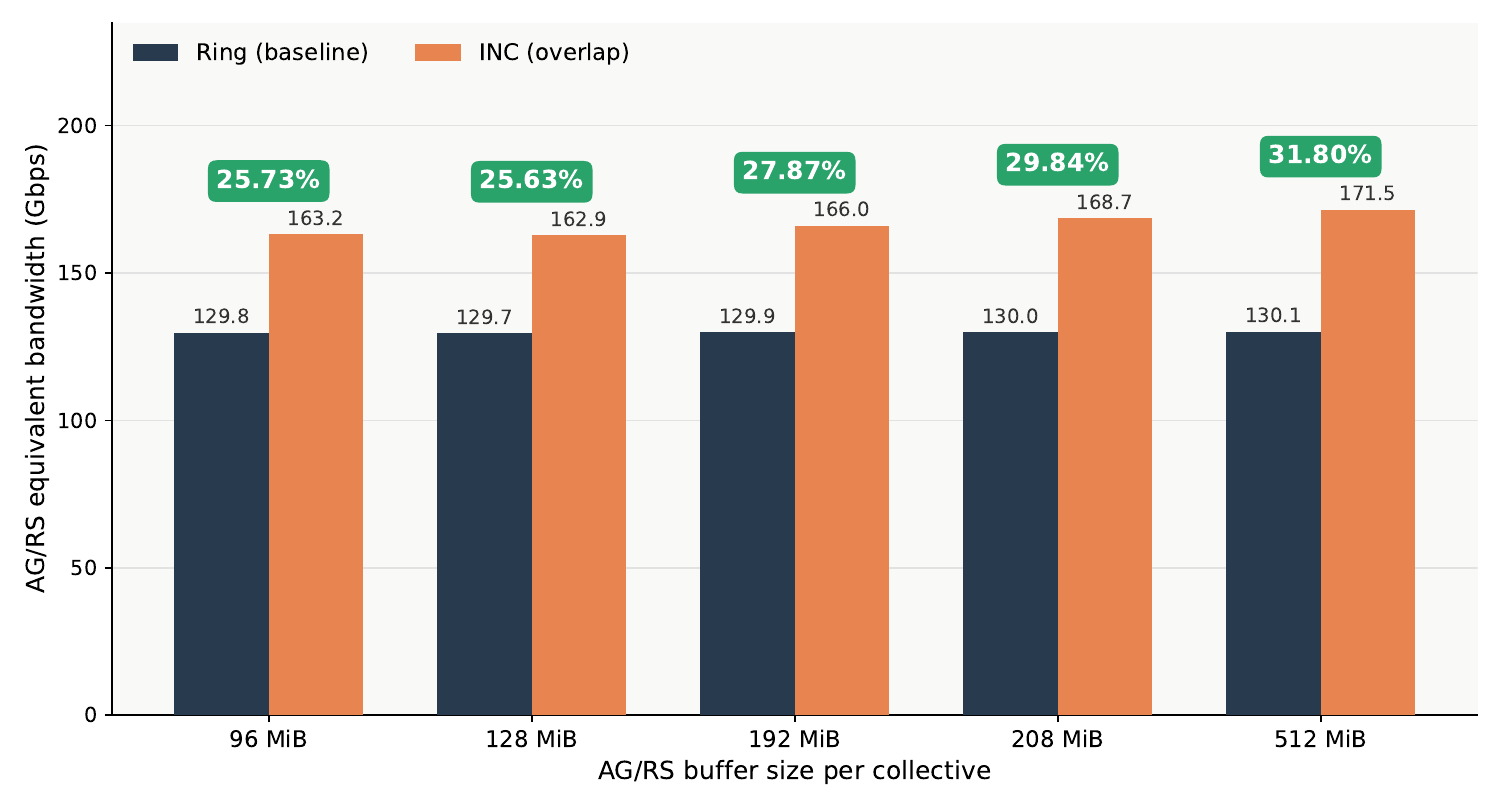}
    \caption{Equivalent bandwidth of concurrently launched \AG{} and \RS{} for increasing buffer size $S$ per collective, across $N{=}4$ ranks at 100\,Gbps per rank: ring collectives executed by the end hosts (baseline) and the in-network collectives of \sys{} (overlap). Equivalent bandwidth is the combined size of the two buffers, $2S$, divided by the time until both collectives complete. Percentages give the bandwidth increase over the baseline, which ranges from 25.6\% to 31.8\%.}
    \label{fig:inc-bw}
\end{figure}

\Cref{fig:inc-bw} shows the results. The ring baseline is consistently flat at about 130\,Gbps for all message sizes, i.e., at 97.5\% of its bound. The overlapped in-network collectives reach 163--172\,Gbps, 25.6--31.8\% more than the ring, and 82--86\% of their bound. Unlike the ring, their equivalent bandwidth increases with the message size.
This is an artifact of the prototype. Because the switch is emulated by an accelerator, every message incurs fixed costs that a hardware switch does not have. For the \RS{}, the emulator serves the $N$ rooted reductions of a message one after another, and each incurs a grid-wide synchronization between the emulator's threads and the fill and drain of its chunk pipeline. Small messages amortize these costs over fewer chunks, which lowers the emulator's efficiency.
A hardware INC switch processes packets at line rate and can serve all roots concurrently, so we expect this dependence on message size, and most of the gap to the bound, to be largely reduced on switch hardware, as they are specific to the emulator rather than to the design.

\subsection{End-to-End Performance}
\label{sec:eval-e2e}
\begin{figure}[ht!]
    \centering
    \includegraphics[width=0.7\linewidth]{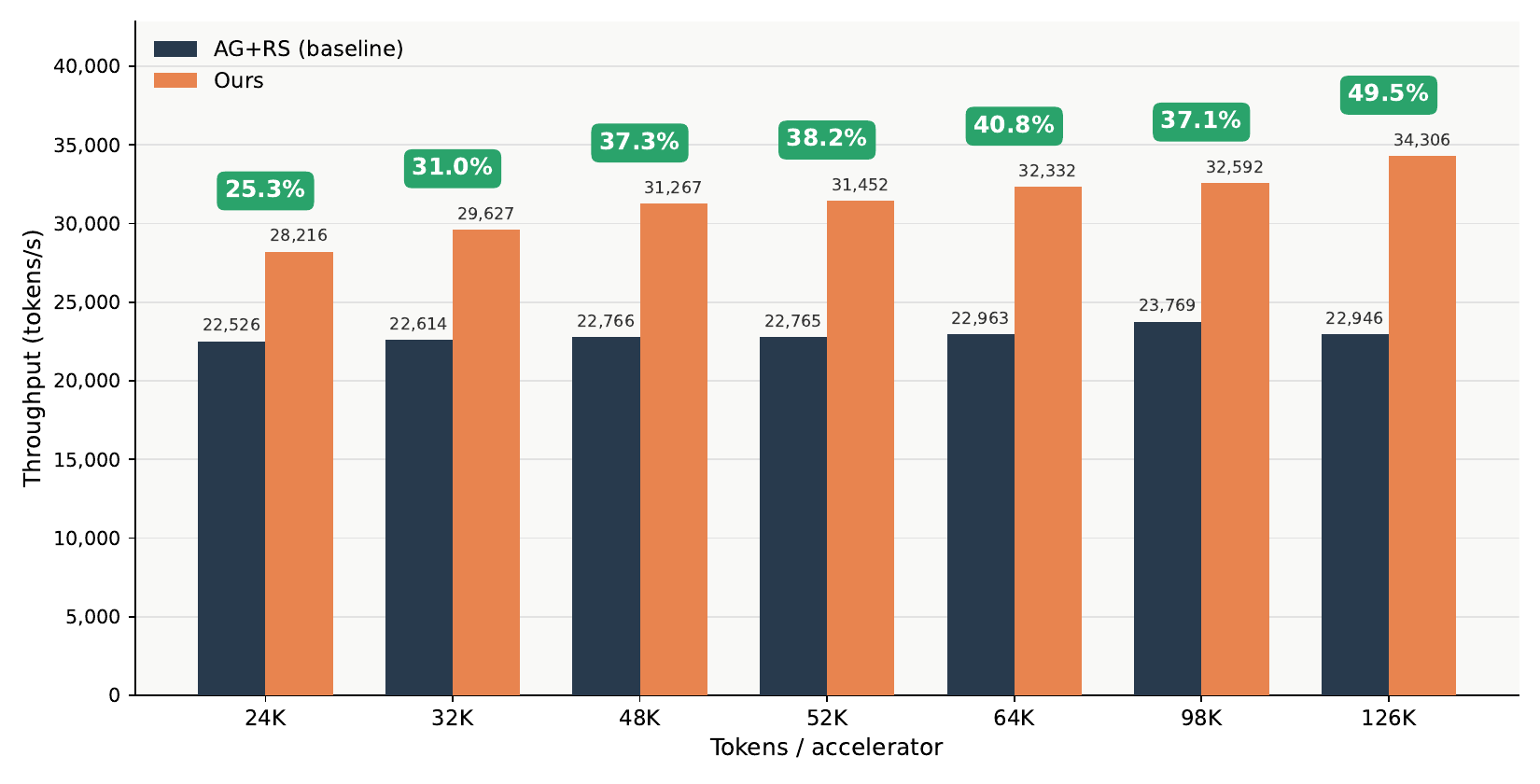}
    \caption{Prefill throughput per accelerator for the same runs as \cref{fig:e2e-ttft}. The baseline's throughput stays
    roughly constant, at about 22.5K--23.8K tokens/s across all prompt sizes, so its
    cost scales linearly with the number of tokens. \sys{} sustains 28.2K--34.3K tokens/s, and its throughput
    increases with prompt size. Percentages give the throughput increase over the baseline,
    $\mathit{tps}_{\mathrm{ours}}/\mathit{tps}_{\mathrm{base}} - 1$, which ranges from 25.3\% to 49.5\%.}
    \label{fig:e2e-tps}
\end{figure}

\Cref{fig:e2e-ttft,fig:e2e-tps} show the TTFT and prefill throughput of \sys{} and the baseline as a function of the number of prompt tokens per accelerator. Both figures derive from the same runs, as throughput is the number of prompt tokens divided by TTFT. \sys{} reduces TTFT by 20.2--33.1\%, which corresponds to a 25.3--49.5\% increase in throughput.

\paragraph{Scaling with prompt size.} The baseline's throughput is nearly constant, at 22.5K--23.8K tokens/s, so its prefill time grows linearly with the number of tokens. The throughput of \sys{} instead rises from 28.2K to 34.3K tokens/s, so its advantage generally grows with the prompt size. Two effects contribute.
First, the full gathered-buffer size per band of the in-network collectives grows with the number of tokens, from 128\,MiB per band at 24K tokens (3 bands) to about 202\,MiB at 126K (10 bands), and the emulated switch processes larger messages slightly more efficiently (\cref{sec:eval-bw}).
Second, as the number of bands grows with prompt size (e.g., from 3 to 10 bands as the number of prompt tokens grows from 24K to 126K), the fill and drain phases, in which the AG and RS do not overlap, take a smaller share of each layer.

\section{Related Work}
\label{sec:related}

\paragraph{MoE systems and EP communication.} GShard~\cite{lepikhin2021gshard}, Switch Transformers~\cite{fedus2022switch}, DeepSpeed-MoE~\cite{rajbhandari2022deepspeedmoe} and Tutel~\cite{hwang2023tutel} established the systems foundations of expert parallelism. For inference, DeepEP~\cite{deepep2025} and pplx-kernels~\cite{pplx2025} provide high-throughput and low-latency all-to-all dispatch and combine kernels, and MegaScale-Infer~\cite{zhu2025megascaleinfer} disaggregates attention and expert computation across devices. These systems optimize communication executed by the end hosts. \sys{} instead offloads it to the switch and uses the \AG{}/\RS{} formulation of dispatch and combine~\cite{tdc2026agrs}.

\paragraph{Overlapping communication and computation.} Flux~\cite{chang2024flux} fuses communication into GEMM tiles for tensor-parallel layers. Comet~\cite{zhang2025comet} overlaps MoE dispatch and combine with expert GEMMs at fine granularity through dependency analysis and adaptive workload assignment. TileLink~\cite{zheng2025tilelink} and Triton-distributed~\cite{zheng2025tritondist} provide tile-level primitives and compiler support for writing overlapped kernels. FlashMoE~\cite{aimuyo2025flashmoe} fuses the entire distributed MoE layer into a single persistent kernel with device-initiated one-sided transfers. At the system level, vLLM's dual batch overlap~\cite{vllmdbo} and SGLang's two-batch overlap~\cite{zheng2024sglang} split a batch into micro-batches so that one micro-batch's communication hides behind the other's computation. Like these works, \sys{} splits tokens into slices that are processed in a pipeline. The split is confined to the MoE layer and runs inside a single kernel with role-specialized blocks. The communication it overlaps is executed by the switch, and it overlaps the two collectives with each other as well as with computation. Micro-batch overlap is orthogonal to \sys{} and could be applied on top of it.

\paragraph{In-network computing.} SHArP~\cite{graham2016sharp} offloads reductions into switch hardware for HPC collectives, and SwitchML~\cite{sapio2021switchml} and ATP~\cite{lao2021atp} perform gradient aggregation on programmable switches for distributed training. The DeepSeek-V3 hardware study~\cite{zhao2025deepseekhw} argues for network hardware with native support for MoE communication patterns.
Closest to our work, DySHARP~\cite{zhang2026dysharp} observes that existing in-switch collectives target static, regular patterns and proposes switch extensions for dynamic multicast and reduction, together with kernel fusion of the dispatch--compute--combine pipeline.
MoE-Hub~\cite{zhou2026moehub} places lightweight hardware in the GPU hub, the switch between a GPU's on-chip crossbar and its scale-up links, to manage address allocation and data flow, so that producers can send data as soon as routing completes. Both propose new hardware for the scale-up fabric. \sys{} takes a complementary position. It operates in the scale-out network with static \AG{} and \RS{} collectives, whose redundancy is small for top-$k$ routing with large $k$ relative to the EP degree (\cref{sec:bg-agrs}), and it focuses on the message-semantic collectives, their concurrent use, and the end-host pipeline needed to turn switch offload into end-to-end gains.

\section{Conclusion}
\label{sec:conclusion}

We presented \sys{}, an expert-parallel MoE layer for the scale-out setting that executes dispatch and combine as in-network AllGather and ReduceScatter and overlaps them with each other and with expert computation in a single band-pipelined kernel.
We implemented both collectives with message semantics using a device-side communication library and an accelerator-based emulator of an INC-capable network switch.
We also showed that, because they load opposite directions of each link, overlapping both collectives reduces the bandwidth-bound communication time by $\frac{N-2}{2(N-1)}$ relative to ring collectives.
We integrated the layer into vLLM as a drop-in replacement.
On OLMoE prefill with four-way EP in an emulated four-node scale-out configuration, \sys{} reduces TTFT by up to 33\% and increases throughput by up to 50\% relative to vLLM's default collective path.
Our results show that switch offload and kernel-level overlap are complementary, and that combined they turn the communication bottleneck of expert parallelism into an opportunity for the network.

\bibliographystyle{plainnat}
\bibliography{references}

\end{document}